\documentclass[10pt,twocolumn]{article}
\usepackage[T1]{fontenc}
\usepackage[utf8]{inputenc}
\usepackage{lmodern}
\usepackage{microtype}
\usepackage{amsmath,amssymb}
\usepackage{graphicx}
\usepackage{booktabs}
\usepackage{cite}
\usepackage{xcolor}
\usepackage{hyperref}
\usepackage[a4paper,margin=1.65cm,columnsep=0.65cm]{geometry}

\hypersetup{colorlinks=true,linkcolor=black,citecolor=black,urlcolor=blue}
\newcommand{\FSL}{T_{\mathrm{FSL}}}
\newcommand{\Ith}{I_{\mathrm{th}}}
\newcommand{\vth}{v_{\mathrm{th}}}
\newcommand{\vref}{v_{\mathrm{ref}}}

\title{A Planar Switching Neuron Shares Logarithmic First-Spike Latency with Its Three-Dimensional Parent But Not Its Onset Mechanism}

\author{%
Giorgio Sonnino$^1$, Vadim Shlyonskiy$^1$, Thomas Erneux$^2$, and David Gall$^1$\\[-0.1em]
\small $^1$Laboratoire d'Enseignement de la Physique, Facult\'e de M\'edecine,\\[-0.2em]
\small Universit\'e libre de Bruxelles, Brussels, Belgium\\[-0.2em]
\small $^2$D\'epartement de Physique des Syst\`emes Dynamiques, Facult\'e des Sciences,\\[-0.2em]
\small Universit\'e libre de Bruxelles, Brussels, Belgium\\[0.2em]
\small Corresponding authors: \href{mailto:giorgio.sonnino@ulb.be}{giorgio.sonnino@ulb.be}, \href{mailto:vadim.shlyonskiy@ulb.be}{vadim.shlyonskiy@ulb.be} }
\date{}

\begin{document}
\maketitle

\begin{abstract}
Can a reduced neuron model reproduce first-spike latency scaling while changing how a stimulus initiates sustained firing? We compare a piecewise-linear planar Neurosimilator with its three-dimensional parent. Both exhibit logarithmic latency dependence over the investigated input intervals, with different fitted coefficients. The mechanisms differ. In the parent, sustained spiking and a stable resting focus coexist in a spiking--rest bistability regime. An abrupt step accesses spiking, while slow stimulation can retain rest at the same final input. Its step-induced firing threshold is therefore an operational basin-access threshold below the saddle-node limit point, not a loss of equilibrium stability. In the planar model, the resting equilibrium instead reaches a switching boundary while its subthreshold eigenvalues remain negative. We derive the planar latency law from this boundary approach. The model generates repetitive excursions without an imposed reset and admits exact trajectories within six switching regions. Similar measured latency scaling thus coexists with different mechanisms of access to firing. Validating a reduced model's mechanism of excitability requires examination of stimulus protocol and attractor access alongside equilibrium structure and spike timing.

\end{abstract}


\section{Introduction}

Low-dimensional neuron models help us understand how neurons cross a firing threshold, generate spikes, and recover. Their value depends on what they retain. Classical models such as the FitzHugh--Nagumo and McKean systems simplify neuronal dynamics while preserving selected features of excitability \cite{FitzHugh1961,McKean1970}. Yet reproducing a voltage trace does not establish that a model preserves the mechanism behind it. This distinction matters when we use reduced models to explain neuronal responses as well as predict them \cite{Chialva2023}. First-spike latency provides a concrete example. Two models may show a similar dependence of latency on input even though different processes govern access to sustained firing. A firing threshold measured with abrupt steps need not mark a bifurcation. In a spiking--rest bistability regime, a stable resting equilibrium and a sustained-spiking attractor coexist at the same input. A change in stimulus protocol can therefore select a different attractor without destabilizing rest. Latency must therefore be interpreted together with the preparation protocol, attractor access, and equilibrium stability. Electronic neurons provide a useful setting for this comparison because their circuit elements implement identifiable feedback pathways. These pathways represent neuronal functions rather than individual ion channels. Piecewise-linear formulations also allow exact trajectory calculations between switching events, making the stages of spike generation accessible to analysis \cite{Tonnelier2003,Coombes2012}. Such models complement broader efforts to construct artificial neurons with diverse firing responses \cite{Yi2018,Feng2025,Shao2025}. Here, we compare the three-dimensional Neurosimilator \cite{Shlyonsky2024} with the two-variable switching model introduced in the companion study \cite{Vadim2026}. The parent combines smooth regenerative feedback with voltage-controlled switching. The planar model replaces its fast depolarizing state with an instantaneous threshold-activated term. This is a heuristic reformulation, not an asymptotic reduction of the parent equations or a reduction of Hodgkin--Huxley channel kinetics \cite{Hodgkin1952,Chialva2023}. We ask whether it retains the measured latency dependence while changing how the stimulus initiates firing. The distinction between two input thresholds is essential. In the parent, $I_{\mathrm{th},\mathrm{3D}}=0.20874~\mathrm{V\,s^{-1}}$ is the step-induced firing threshold for the preparation protocol used here. For inputs in the studied interval above this basin-access threshold, an abrupt step can initiate sustained spiking while a sufficiently slow ramp to the same final input leads to accommodation at a stable resting focus. The latency measurements concern a short input interval in this spiking--rest bistability regime, below the equilibrium limit point $I_{\mathrm{LP}}\simeq0.296427~\mathrm{V\,s^{-1}}$. They do not probe firing onset at the fold. In the planar model, the reference threshold instead marks the loss of admissibility of the subthreshold equilibrium at a switching boundary. We compare voltage excursions, the full-range frequency--current response, and step-induced first-spike latency. Both models exhibit logarithmic latency dependence over the investigated intervals, with different fitted coefficients. We derive the planar law from the approach to the switching boundary; we do not infer a corresponding mechanism or an asymptotic law for the parent from its finite-range fit. The saddle-node-on-invariant-circle (SNIC) interpretation of the parent's full-range frequency response is a separate issue from access to an existing spiking attractor in the latency experiment. The planar reformulation makes the relation between switching dynamics and spike timing analytically accessible. Comparison with the parent then exposes a limit to what similar latency scaling establishes: it need not preserve protocol-dependent access to firing or spiking--rest bistability. Validation of a reduced neuronal model must therefore address the stimulus history and attractors reached, as well as the measured timing response and local equilibrium structure.

\section{Planar switching model}

The reduced model tracks a voltage-like variable $v$ and a recovery variable $k$:
\begin{align}
&\!\!\dot v\!=\!I\!-\!Av\!+\!P\Theta(v\!-\!\vth)\!-\!Gk\!-\!E(k-\vth)\Theta(k\!-\!\vth)\label{eq:v}\\
&\!\!\dot k\!=\!Fv_s\Theta(v\!-\!\vref)\!-\!Fk\!+\!F(\vth\!-\!k)\Theta(k\!-\!\vth) \label{eq:k}
\end{align}
Here, $\Theta$ denotes the Heaviside step function and $P=F(v_s-V_0)$, where $V_0=1~\mathrm{V}$ serves as the circuit reference. The input $I$ represents the injected current normalized by membrane capacitance. Because of this scaling, its unit is voltage per unit time rather than current. Throughout this study, we fix the parameters as:
\begin{equation}
\begin{split}
&\vth=0.7~\mathrm{V},\quad \vref=3.3~\mathrm{V},\quad
v_s=7.5~\mathrm{V},\\
&A=0.02~\mathrm{ms}^{-1},\  E=20~\mathrm{ms}^{-1},\ 
F=0.4~\mathrm{ms}^{-1},\\
&G=0.2~\mathrm{ms}^{-1}
\end{split}
\label{eq:parameters}
\end{equation}
These mathematical terms map directly to circuit-level leakage, threshold-activated depolarization, recovery, and repolarization; they are not intended to represent specific physiological sodium or potassium currents. Crucially, Equations~\eqref{eq:v} and \eqref{eq:k} do not rely on an imposed reset mechanism. Instead, the piecewise-defined dynamics route the trajectory through a large depolarizing excursion, followed by repolarization and subsequent recovery. The boundaries defined by $v=\vth$, $v=\vref$, and $k=\vth$ slice the phase plane into six distinct linear regions, where a standard firing cycle steps sequentially through:
\begin{equation}
R_1\to R_2\to R_3\to R_4\to R_5\to R_6\to R_1
\label{eq:sequence}
\end{equation}

\section{What the reduction preserves}

\begin{figure*}[t]
\centering
\includegraphics[width=0.96\textwidth]{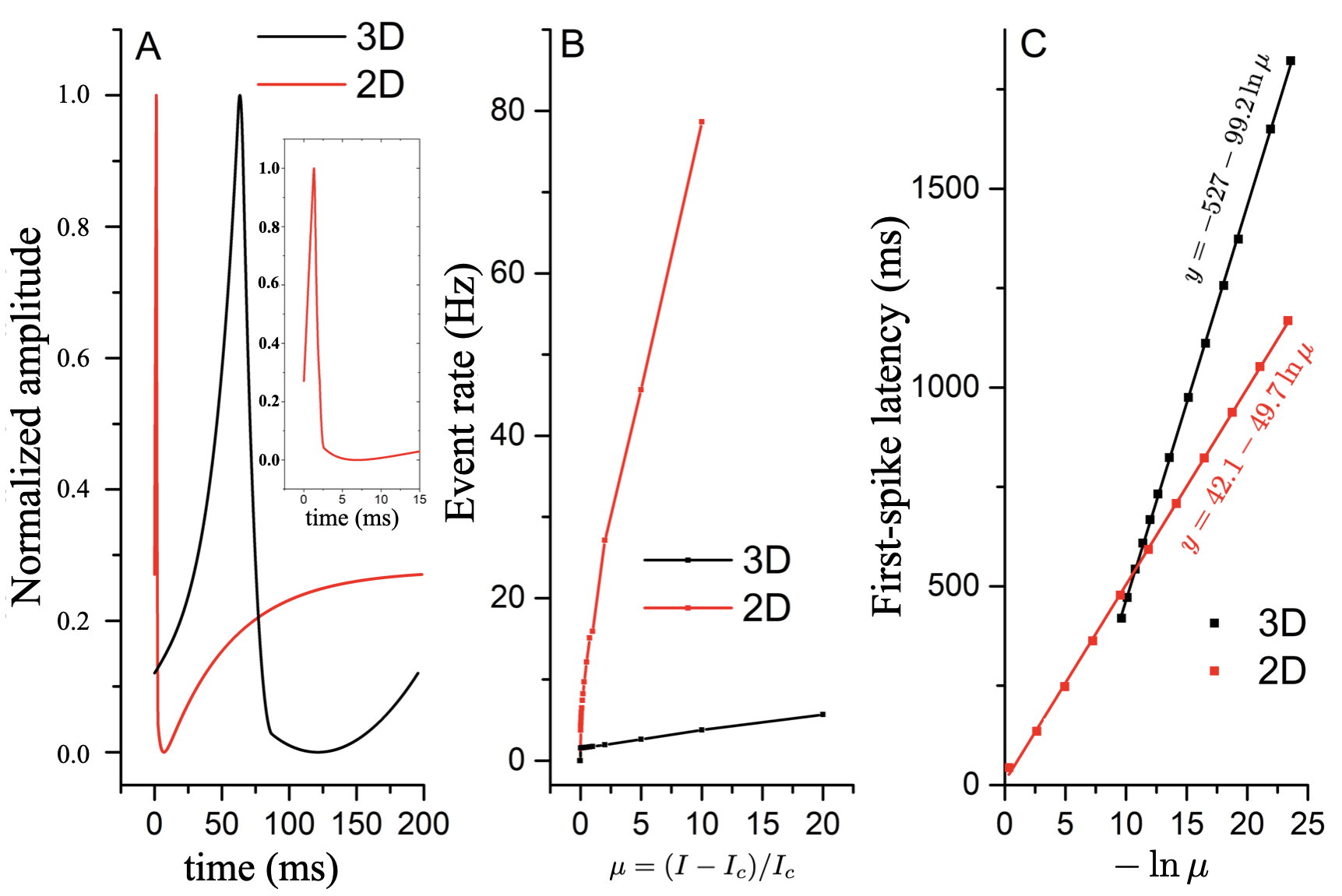}
\caption{\textbf{Comparison of the 3D and 2D models.} (A) Normalized voltage amplitude versus time. The inset resolves the sharp 2D excursion over $0$--$15~\mathrm{ms}$. The 3D trace reaches its maximum later and follows a different recovery path. (B) Event rate over the full input range shown, plotted against $\mu=(I-I_c)/I_c$ as defined in Eq.~\eqref{eq:mu}. The reference inputs are $I_{c,\mathrm{3D}}=0.20874~\mathrm{V\,s^{-1}}$ and $I_{c,\mathrm{2D}}=0.014~\mathrm{V\,ms^{-1}}$, not the parent's fold input. The 2D response has a much larger gain over the plotted range. (C) Step-induced first-spike latency versus $-\ln\mu$. The parent data concern a short interval above its basin-access threshold and below $I_{\mathrm{LP}}$, within the spiking--rest bistability regime. They are not a near-fold latency measurement. The straight lines are $\FSL^{\mathrm{3D}}\simeq(-527-99.2\ln\mu)~\mathrm{ms}$ and $\FSL^{\mathrm{2D}}\simeq(42.1-49.7\ln\mu)~\mathrm{ms}$ over the intervals shown. Black and red denote the 3D and 2D models, respectively.}
\label{fig:comparison}
\end{figure*}

A direct comparison between the parent 3D model~\cite{Shlyonsky2024} and the planar model introduced in the companion letter~\cite{Vadim2026} is illustrated in Figure~\ref{fig:comparison}. Because the individual voltage traces in panel (A) are normalized separately, this panel evaluates waveform morphology and timing rather than physical amplitude. While both traces display a pronounced excursion followed by a recovery phase, the similarities end there; the 2D event is visibly sharper, and their recovery trajectories operate on distinct timescales. To establish a dimensionless metric for the applied drive, we define the relative distance-from-threshold as:
\begin{equation}
\mu_j=\frac{I_j-I_{c,j}}{I_{c,j}},\qquad j\in\{\mathrm{3D},\mathrm{2D}\}
\label{eq:mu}
\end{equation}
where $I_{c,\mathrm{3D}}=I_{\mathrm{th},\mathrm{3D}}=0.20874~\mathrm{V\,s^{-1}}$ is the step-induced basin-access threshold, and $I_{c,\mathrm{2D}}=I_{\mathrm{th},\mathrm{2D}}=A\vth=0.014~\mathrm{V\,ms^{-1}}$ is the planar switching threshold. In particular, $I_{c,\mathrm{3D}}\ne I_{\mathrm{LP}}$. Equation~\eqref{eq:mu} expresses distance from each model's stated reference threshold; it does not identify these thresholds with the same dynamical event. The native input units differ between the models. Panel (B) plots the event rate under constant input over the full input range displayed, rather than only the narrow spiking--rest bistability interval used for latency. Discussion of the parent's SNIC and limit-point structure belongs to this broader frequency--current response; it must not be transferred to the step-induced threshold used in the latency experiment. Although both rates scale upward with $\mu$, their gains diverge by more than an order of magnitude across the illustrated ranges. Specifically, the 2D model approaches $80~\mathrm{Hz}$ while the 3D model stays below roughly $7~\mathrm{Hz}$. Ultimately, the reduction captures the directional trend of the input-output response, rather than its exact quantitative gain.
Panel (C) shows the first-spike latency, which is distinct from the mean interspike interval during repetitive firing. Assuming a constant input step initiates at $t=t_{\mathrm{step}}$ and the initial upward threshold crossing happens at $t=t_1$, the latency is given by:
\begin{equation}
\FSL=t_1-t_{\mathrm{step}}
\label{eq:fsl}
\end{equation}
Isolating the first spike is important in the parent. The initial step accesses the spiking attractor from the prepared resting state, whereas subsequent spikes follow the established oscillatory trajectory. Their transit times need not agree. Averaging the first latency with subsequent interspike intervals would mix distinct observables. For the parent, the measured latency is referenced to the operational threshold $I_{\mathrm{th},\mathrm{3D}}=0.20874~\mathrm{V\,s^{-1}}$, not the equilibrium limit point.\footnote{Ref.~\cite{Shlyonsky2024} described the repetitive spiking period using an inverse-square-root law referenced to the analytically derived limit point, $I_{\mathrm{LP}}\simeq0.296~\mathrm{V\,s^{-1}}$. The present first-spike latency analysis instead uses $\mu_{\mathrm{3D}}=(I-I_{\mathrm{th},\mathrm{3D}})/I_{\mathrm{th},\mathrm{3D}}$, with $I_{\mathrm{th},\mathrm{3D}}=0.20874~\mathrm{V\,s^{-1}}$, in the protocol-dependent spiking--rest bistability regime below the fold. In the finite-range fit comparison, the inverse-square-root representation gives $R^2\simeq0.95$, while the logarithmic representation gives $R^2\geq0.99$. The logarithmic representation is therefore preferred empirically over the resolved interval; goodness of fit alone neither excludes the alternative nor identifies a bifurcation mechanism. A first-spike latency fit at the basin-access threshold cannot by itself overturn an interpretation of the repetitive period or full-range frequency response near the SNIC.}
At inputs in the studied spiking--rest bistability interval, an abrupt step initiates sustained firing, whereas a slow ramp to the same final input allows accommodation towards a stable focus. We use \emph{basin-access threshold} to describe the lowest step amplitude that accesses sustained spiking for this preparation protocol. Its value is specific to the baseline state and step protocol used in the latency experiment. It is not a universal boundary for the existence of the oscillatory attractor and need not coincide with a saddle-node or Hopf bifurcation. Across the resolved interval, the parent latency data vary approximately linearly with $-\ln\mu_{\mathrm{3D}}$. The fitted coefficients are approximately $99.2~\mathrm{ms}$ in the parent and $49.7~\mathrm{ms}$ in the planar model. Similar functional form does not imply equal latency or a common mechanism. In particular, we do not interpret the parent fit as passage through a saddle-node bottleneck at $I_{\mathrm{LP}}$, since the latency experiment concerns a different operational regime.

\section{Similar latency scaling with different access to firing}

\begin{figure}[t]
\centering
\includegraphics[width=\columnwidth]{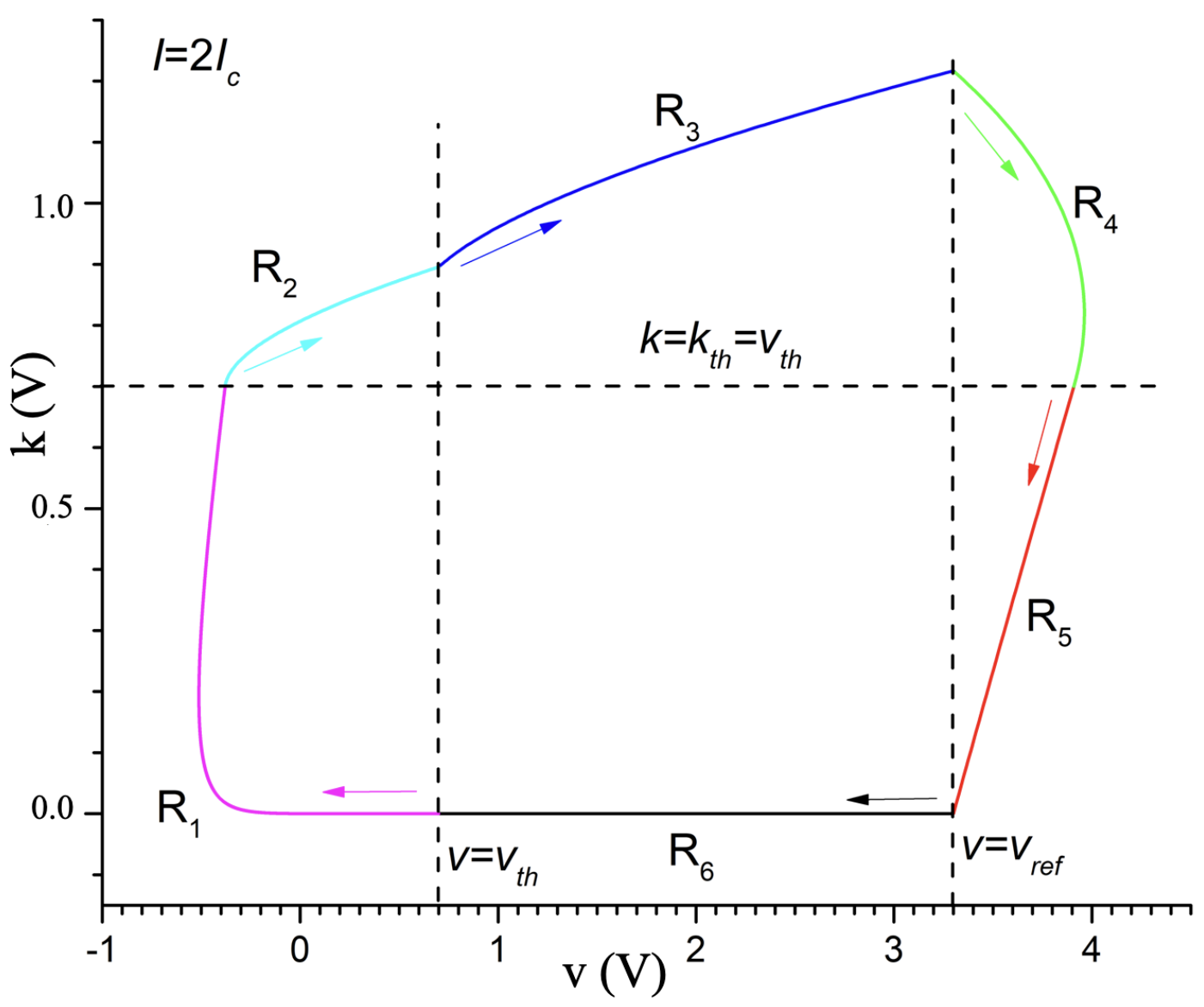}
\caption{\textbf{Planar firing cycle through the six switching regions.} At $I=2I_{c,\mathrm{2D}}$, the orbit follows Eq.~\eqref{eq:sequence}; arrows indicate motion. Dashed lines mark $v=\vth$, $v=\vref$, and $k=\vth$. Boundary crossings were located with event-resolved DOP853 integration; plotting uses a separate sampling grid.}
\label{fig:phase}
\end{figure}

\begin{figure}[!ht]
\centering
\includegraphics[width=\columnwidth]{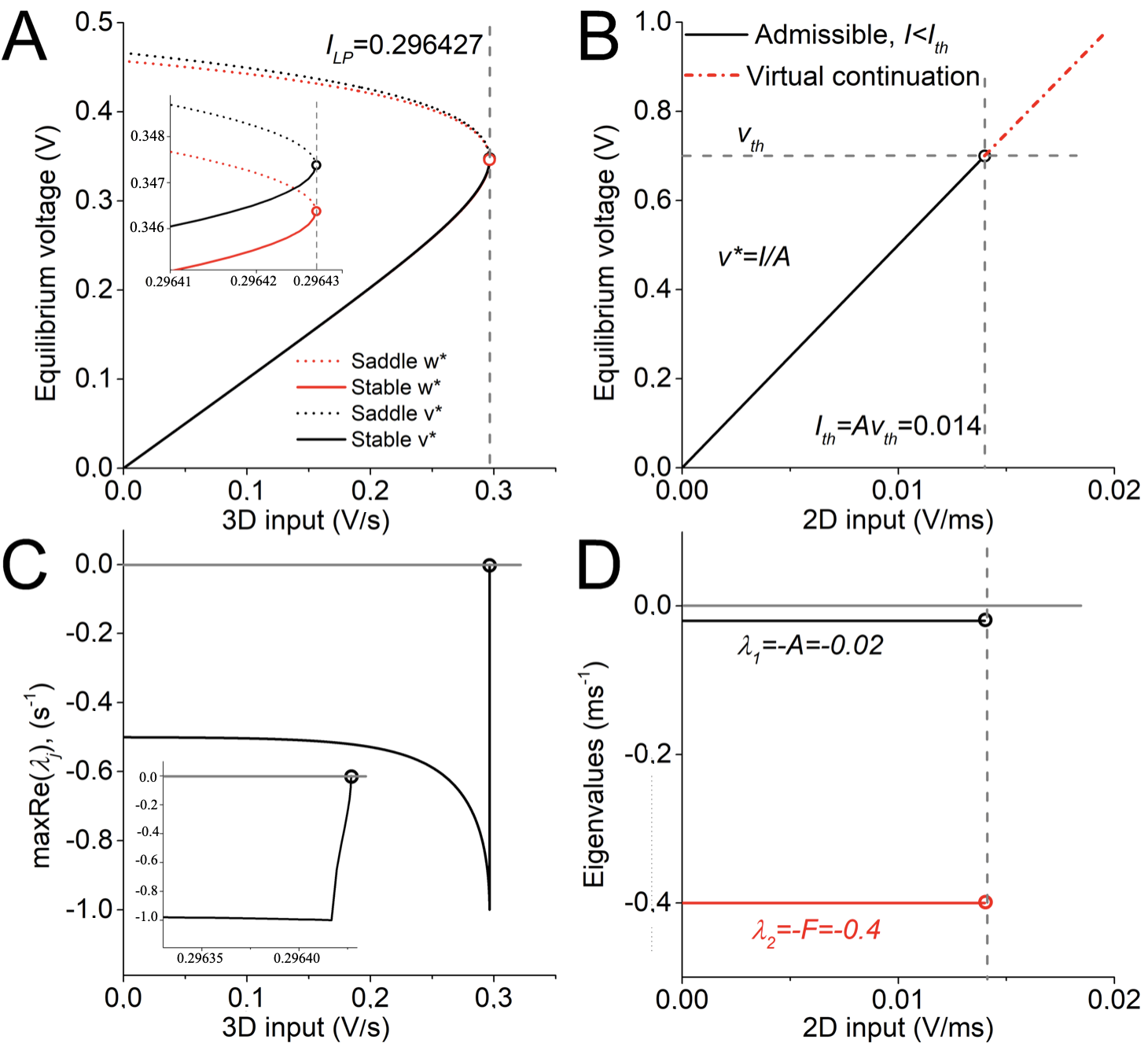}
\caption{\textbf{Equilibrium structure and the distinction between basin access and the fold.}
(A) Parent equilibrium coordinates $v^*$ (black) and $w^*$ (red): solid curves denote the stable branch and dotted curves the saddle branch. The dashed line marks $I_{\mathrm{LP}}=0.296427~\mathrm{V\,s^{-1}}$; the inset enlarges its neighbourhood. This fold is distinct from the step-induced threshold $I_{\mathrm{th},\mathrm{3D}}=0.20874~\mathrm{V\,s^{-1}}$ used in Figure~\ref{fig:comparison}(C).
(B) The planar equilibrium $v^*=I/A$, $k^*=0$, reaches its admissibility boundary at $\Ith=A\vth=0.014~\mathrm{V\,ms^{-1}}$. The red dotted segment is a virtual continuation, not an equilibrium of the full switched system.
(C) Largest real part of the parent equilibrium eigenvalues. The inset resolves the approach to zero near the fold; the open circle marks the limit point, not the basin-access threshold.
(D) Planar subthreshold eigenvalues $-A=-0.02~\mathrm{ms^{-1}}$ and $-F=-0.4~\mathrm{ms^{-1}}$. Open endpoints indicate limits from below, not Jacobians on the switching surface.
Input and eigenvalue units are $\mathrm{V\,s^{-1}}$ and $\mathrm{s^{-1}}$ in (A,C), and $\mathrm{V\,ms^{-1}}$ and $\mathrm{ms^{-1}}$ in (B,D).}
\label{fig2a}
\end{figure}

In the planar subthreshold region, the switching terms vanish, and the resting equilibrium is
\begin{equation}
(v^*,k^*)=(I/A,0)
\label{eq:eqpoint}
\end{equation}
Its eigenvalues remain $-A$ and $-F$ as input increases. At
\begin{equation}
I=\Ith=A\vth=0.014~\mathrm{V\,ms^{-1}}
\label{eq:ith}
\end{equation}
the stable node reaches $v=\vth$ and ceases to be admissible. The continuation $v^*=I/A$ then lies outside the subthreshold region: no equilibrium pair collides and neither eigenvalue vanishes. Figure~\ref{fig:phase} shows the resulting switching cycle.
During the long subthreshold transit in $R_1$, $k$ relaxes towards zero, reducing the voltage equation to
\begin{equation}
\dot v=I-Av
\end{equation}
Starting from $v_0$, this approximation gives the threshold-crossing time
\begin{equation}
t_1=\frac{1}{A}\ln\!\left(\frac{I-Av_0}{I-A\vth}\right)
\label{eq:t1}
\end{equation}
For $v_0$ a fixed distance below $\vth$, as $I\to\Ith^+$,
\begin{equation}
t_1=-\frac{1}{A}\ln\mu_{\mathrm{2D}}+O(1)
\label{eq:loglaw}
\end{equation}
The predicted coefficient, $1/A=50~\mathrm{ms}$, agrees closely with the fitted value of $49.7~\mathrm{ms}$ in Figure~\ref{fig:comparison}(C). The parent's similar empirical latency dependence concerns a different mechanism. Its step-induced basin-access threshold, $I_{\mathrm{th},\mathrm{3D}}=0.20874~\mathrm{V\,s^{-1}}$, lies in a spiking--rest bistability regime: an abrupt step accesses sustained spiking, whereas a slow ramp to the same input allows accommodation at a stable focus. Figure~\ref{fig2a} places this regime below the equilibrium fold at $I_{\mathrm{LP}}\simeq0.296427~\mathrm{V\,s^{-1}}$. The zero eigenvalue at that fold does not explain the latency experiment near the lower basin-access threshold. The relevant contrast is therefore protocol-dependent attractor selection in the parent versus loss of equilibrium admissibility in the planar model. The equilibrium plots provide structural context, not a basin map; the shared logarithmic fits alone do not identify a common mechanism.

\section{Circuit realization, limits, and interpretation}

\begin{figure}[t]
\centering
\includegraphics[width=0.4\textwidth]{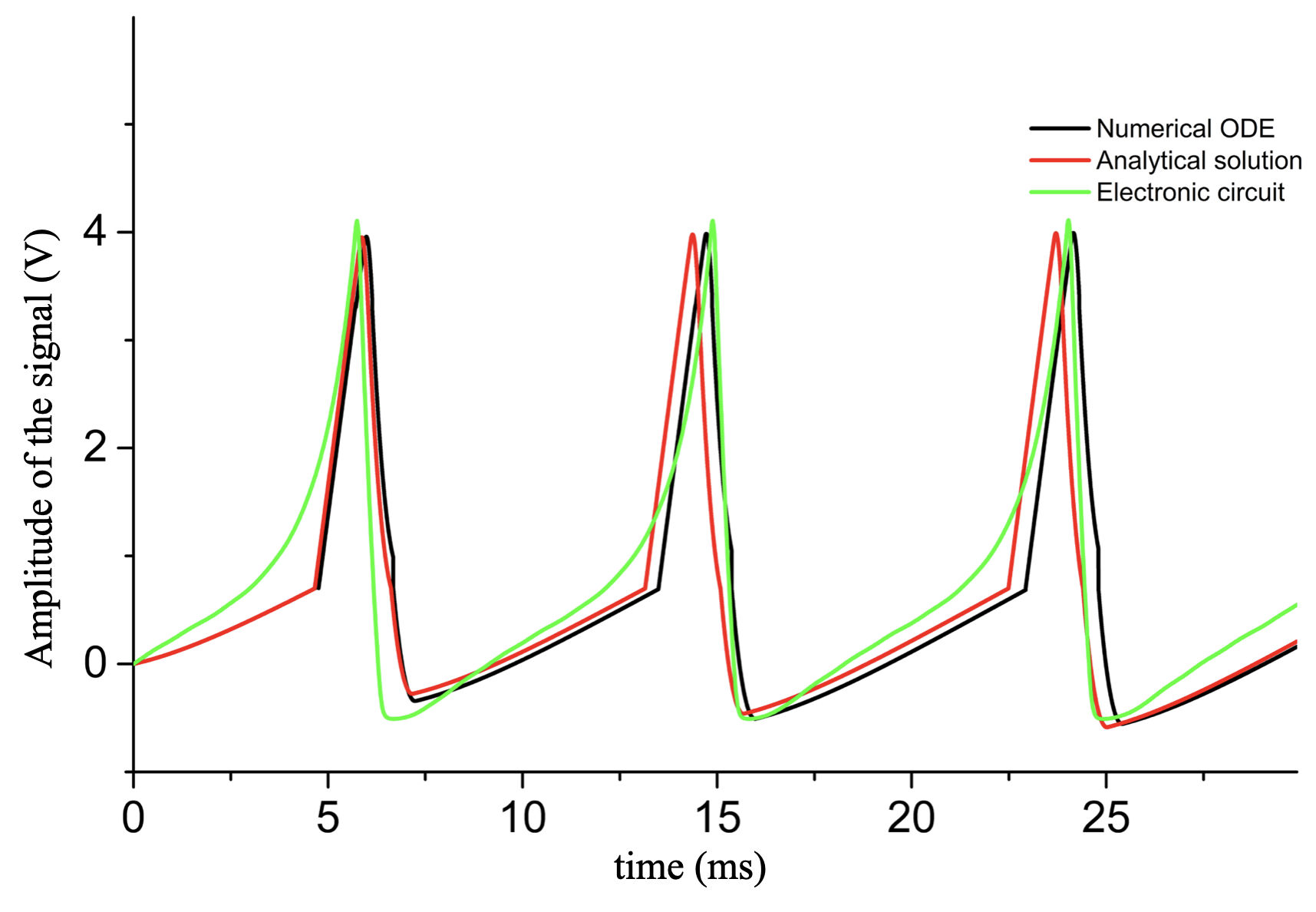}
\caption{\textbf{Piecewise-linear neuron model with analytical and hardware realisation.} This figure compares the numerical solution of the ODEs (in blue), the stepwise closed-form solution (in red), and the measured signal from the electronic implementation documented in Supplementary Figure~\ref{fig:pcb}. This comparison concerns agreement between the planar model and its implementation; it does not establish equivalence with the three-dimensional parent or biological neurons. These results were obtained at much higher input levels, which accounts for the difference in spike shape compared with Fig.~\ref{fig:comparison}A}
\label{fig5}
\end{figure}

The planar dynamical equations map directly onto the compact hardware platform, illustrated in Supplementary Figure~\ref{fig:pcb}. Threshold tracking and recovery mechanisms are handled by discrete switching and comparator stages, with the entire printed circuit board powered via a standard USB-C interface. While this physical implementation demonstrates that the two-variable switching framework is readily realizable using off-the-shelf electronic components, it serves as a functional analog; it does not map the underlying mathematical terms onto literal biophysical ion currents. The circuit schematic of the 2D model can be found in the companion manuscript \cite{Vadim2026} and in \cite{Dupuis2024}. The signal provided by the electronic circuit has been compared with the numerical and stepwise closed-form solutions of the ordinary differential equations (ODEs) (see Figure~\ref{fig5}).
\noindent Several scope boundaries constrain the claims made in this work. Foremost, the dimensional reduction is fundamentally heuristic: it is engineered to capture specific input-output response characteristics, but it sacrifices the true 3D bifurcation topology, the quantitative frequency-current gain profile, and the broader spectrum of complex periodic, quasiperiodic, or chaotic behaviors. Furthermore, because the model relies heavily on multiple Heaviside step functions, precise numerical event localization becomes paramount. Utilizing a coarse, fixed integration step introduces localized boundary-crossing shifts that can rapidly degrade the integrity of simulated spike timings. An additional limitation stems from the fact that the piecewise analytical calculations remain exact only within the interior of each linear domain. Approximations made when solving for crossing times inevitably inject phase errors that accumulate over time, an issue that becomes particularly pronounced under high-drive conditions. The parent latency fit is empirical and concerns protocol-dependent access to firing in the spiking--rest bistability regime. It neither locates a bifurcation nor characterizes asymptotic dynamics at the higher-input equilibrium fold. Dropping the third state introduces a puzzle that this comparison leaves open: which specific responses rely on that state's dynamics, and to what extent can a basic history-dependent term win back that influence? Because the remaining recovery variable still holds temporal information, it would be wrong to call the planar model memoryless. Instead, it simply bypasses a specific internal dynamical loop. The preliminary role of $w$ in step-versus-ramp selection motivates studying this omitted history dependence. To illustrate how eliminating a relaxing state can introduce memory, consider a local linear representation, $\dot{\widetilde w}=a\widetilde v-b\widetilde w$ (where $b>0$). Integrating this yields:
\begin{equation} 
\widetilde w(t)=e^{-bt}\widetilde w(0) + a\int_0^t e^{-b(t-s)}\widetilde v(s) ds
\end{equation} 
 This expression retains both the initial internal state and the voltage history. It illustrates the memory introduced by elimination under the stated local linear approximation; it is not an exact closure of the nonlinear parent or a proof of the mechanism of basin selection. Using delayed representations for recovery dynamics is not entirely new; Paraskevov explored a delayed representation in a Morris--Lecar reduction \cite{Paraskevov2022}. This brings up the next obvious question: how well can a basic memory term restore a response that the instantaneous model wipes out? A follow-up study will evaluate accommodation and the sustained firing range across three setups: the parent model, the planar version, and the planar model equipped with linear delayed feedback. This comparison will test whether recovering sensitivity to stimulus history changes the input range supporting sustained firing. For now, this Letter maps out the structural differences that any future functional recovery must be measured against. A two-variable delay model requires a history state and therefore no longer has a strictly two-dimensional phase space.

\section{Conclusion}

The planar Neurosimilator retains large spike-like excursions and an event rate that increases with input, but its waveform timing and frequency--current gain differ substantially from those of its three-dimensional parent. Both models exhibit logarithmic first-spike latency dependence over the investigated intervals, with different fitted coefficients. The shared functional form does not establish a common mechanism of access to firing. In the parent latency experiment, $I_{\mathrm{th},\mathrm{3D}}=0.20874~\mathrm{V\,s^{-1}}$ is a step-induced basin-access threshold in a spiking--rest bistability regime, where sustained spiking and a stable resting focus coexist. An abrupt step initiates firing, while slow stimulation permits accommodation. This threshold is distinct from the higher-input saddle-node limit point and is not a loss of equilibrium stability. In the planar model, the resting equilibrium instead becomes inadmissible at a switching boundary. We derive its logarithmic latency law from the approach to that boundary. The gain from the reformulation is analytical access to the switching stages and a concrete account of what similar timing responses fail to preserve. Validation of a reduced neuronal model must address stimulus preparation and attractor selection alongside spike timing and equilibrium structure. The parent's full-range frequency response and its SNIC interpretation remain distinct from the finite-range basin-access latency experiment. Preliminary observations implicate the eliminated variable $w$ in protocol-dependent selection, but do not yet establish the controlling basin geometry. A subsequent study will test which history-dependent responses a simple memory representation can recover and at what cost to sustained firing.

\section*{CRediT authorship contribution statement}
The authors contributed equally to the work.

\section*{Declaration of competing interest}
The authors declare that they have no known competing financial interests or personal relationships that could have influenced the work reported in this paper.

\section*{Data and code availability}
The code used to generate and analyse the results is available at \url{https://gitlab.ulb.be/vshlyons/2d-neurosimilator}. PCB photos of the 3D model circuit can be seen in \cite{Shlyonsky2024} and \cite{Dupuis2024}.

\clearpage
\onecolumn
\section*{Supplementary material}
\setcounter{figure}{0}
\renewcommand{\thefigure}{S\arabic{figure}}
\renewcommand{\theHfigure}{S\arabic{figure}}
\subsection*{Stagewise solution expressions}

\noindent\textbf{Stage 1}
\begin{equation}
k = k_0 e^{-Ft},\quad v = \left[v_0-\frac{I}{A}+\frac{Gv_{\mathrm{th}}}{A-F}\right]e^{-At} +\frac{I}{A}-\frac{Gv_{\mathrm{th}}}{A-F}e^{-Ft}\nonumber
\end{equation}
\noindent\textbf{Stage 2}
\begin{align*}
k &= k_{t_1}e^{-Ft}, \\
v &= \frac{k_{t_1}G}{F-A}\left(e^{-Ft}-e^{-At}\right)
+\left(v_{\mathrm{th}}-\frac{I+F(v_s-1)}{A}\right)e^{-At}
+\frac{I+F(v_s-1)}{A}.
\end{align*}
\noindent\textbf{Stage 3}
\begin{align*}
k &= v_s+(k_{t_2}-v_s)e^{-Ft}, \\
v &= \left(
v_{\mathrm{ref}}-\frac{I+F(v_s-1)}{A}
+\frac{G(k_{t_2}-v_s)}{A-F}+\frac{Gv_s}{A}
\right)e^{-At} \\
&\quad -\frac{G(k_{t_2}-v_s)}{A-F}e^{-Ft}+\frac{I+F(v_s-1)}{A}-\frac{Gv_s}{A}
\end{align*}
\noindent\textbf{Stage 4}
\begin{align*}
k =\ &\frac{v_{\mathrm{th}}+v_s}{2}
+\frac{v_{\mathrm{th}}-v_s}{2}e^{-2Ft}, \\
v =\ &\frac{(v_s-v_{\mathrm{th}})(G+E)}{2(2F-A)}
\left(e^{-At}-e^{-2Ft}\right) \\
&+\left(
v_{t_3}-\frac{I+F(v_s-1)}{A}-\frac{v_s(G+E)+v_{\mathrm{th}}(G-E)}{2A}
\right)e^{-At} +\frac{I+F(v_s-1)}{A}\\
&-\frac{v_s(G+E)+v_{\mathrm{th}}(G-E)}{2A}
\end{align*}
\noindent\textbf{Stage 5}
\begin{align*}
k &= \left(k_{t_4}-\frac{v_{\mathrm{th}}}{2}\right)e^{-2Ft}
+\frac{v_{\mathrm{th}}}{2}, \\
v &= \left(k_{t_4}-\frac{v_{\mathrm{th}}}{2}\right)
\frac{G+E}{2F-A}\left(e^{-2Ft}-e^{-At}\right) +\left(
v_{\mathrm{ref}}-\frac{I+F(v_s-1)}{A}+\frac{v_{\mathrm{th}}(E-G)}{2A}
\right)e^{-At} \\
&\quad +\frac{I+F(v_s-1)}{A}
+\frac{v_{\mathrm{th}}(E-G)}{2A}
\end{align*}
\noindent\textbf{Stage 6}
\begin{align*}
k &= \left(k_{t_5}-\frac{v_{\mathrm{th}}}{2}\right)e^{-2Ft}
+\frac{v_{\mathrm{th}}}{2}, \\
v &= \left(k_{t_5}-\frac{v_{\mathrm{th}}}{2}\right)
\frac{G+E}{2F-A}\left(e^{-2Ft}-e^{-At}\right)
+v_{\mathrm{th}}e^{-At} +\left(
\frac{I}{A}+\frac{v_{\mathrm{th}}(E-G)}{2A}
\right)\left(1-e^{-At}\right)
\end{align*}

\clearpage
\begin{figure}[!ht]
\centering
\includegraphics[width=0.65\textwidth]{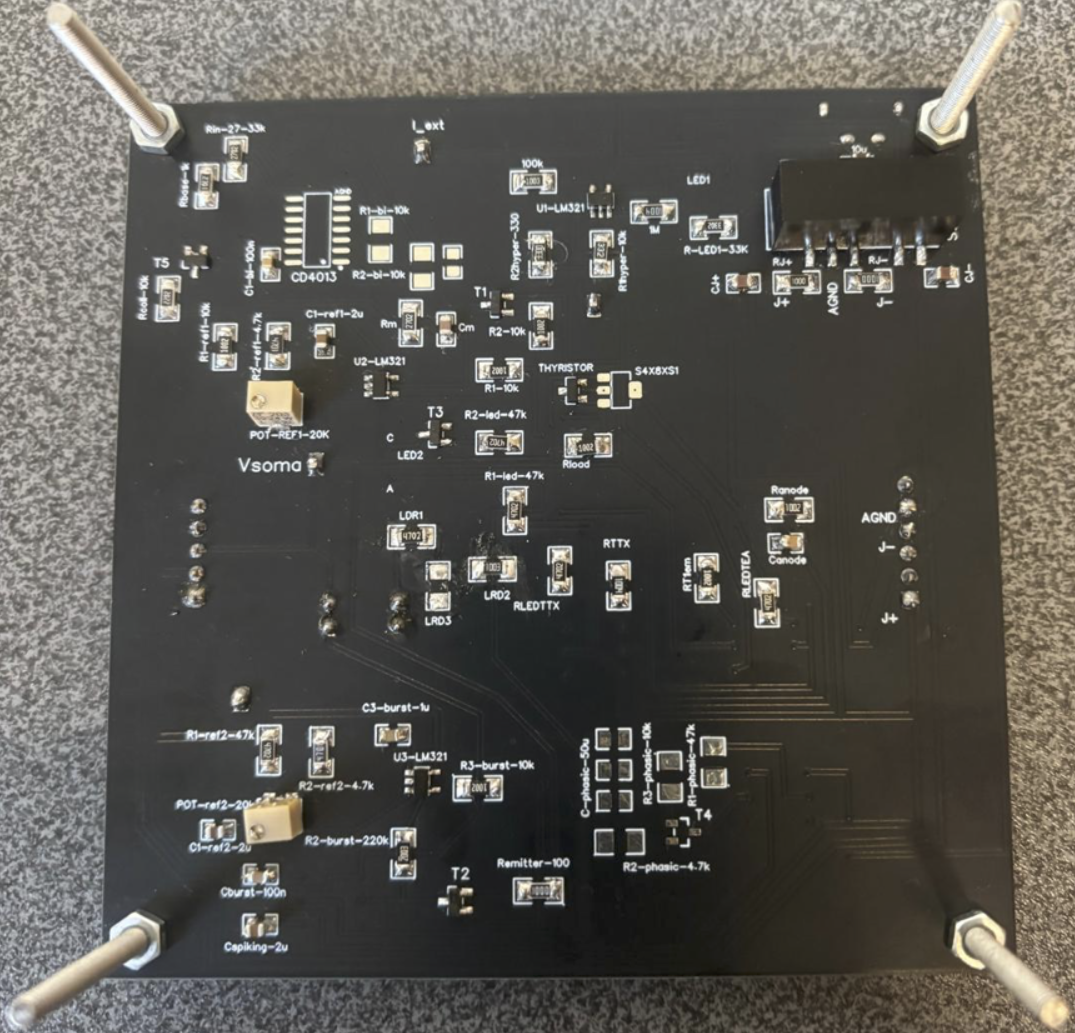}
\caption{\textbf{Electronic implementation of the planar model.} Photograph of the printed circuit board containing the switching and recovery stages. The board uses USB-C power. The photograph documents the hardware implementation; comparison of its measured voltage signal with the model is presented in Figure~\ref{fig5}.}
\label{fig:pcb}
\end{figure}

\end{document}